# Access to Personal Data and the Right to Good Governance during Asylum Procedures after the CJEU's YS. and M. and S. judgment

Evelien Brouwer and Frederik Zuiderveen Borgesius

E.R. Brouwer is associate professor at the section migration law, VU University Amsterdam

F.J. Zuiderveen Borgesius is a researcher at the Institute for Information Law (IViR), University of Amsterdam

*Abstract*

In the *YS. and M. and S.* judgment, the Court of Justice of the European Union ruled on three procedures in which Dutch judges asked for clarification on the right of asylum seekers to have access to the documents regarding the decision on asylum applications. The judgment is relevant for interpreting the concept of personal data and the scope of the right of access under the Data Protection Directive, and the right to good administration in the EU Charter of Fundamental Rights. At first glance, the judgment seems disappointing from the viewpoint of individual rights. Nevertheless, in our view the judgment provides sufficient grounds for effective access rights to the minutes in future asylum cases.

**Table of Contents**

## 1 Introduction

In the *YS. and M. and S.* judgment (17 July 2014), the Court of Justice of the European Union (CJEU) ruled on three procedures in which Dutch judges asked for clarification on the right of asylum seekers to have access to the minutes underlying the decision on asylum applications.[1] These questions address a specific feature of the Dutch asylum procedure. But the CJEU's reasoning may be relevant for interpreting the concept of 'personal data' and the scope of the right of access under the Data Protection Directive,[2] and the right to good administration in Article 41 (1) of the EU Charter of Fundamental Rights.

At first glance, the judgment seems disappointing from the viewpoint of individual rights, because the CJEU seems to interpret 'personal data' and the access right restrictively, and refuses to apply Article 41 of the Charter. Nevertheless, in our view the judgment provides sufficient grounds for effective access rights to the minutes in future asylum cases.

The article is structured as follows. In section 2 we briefly discuss the significance of minutes in Dutch asylum procedures. In section 3 we summarise the *YS. and M. and S.* judgement. In section 4 we comment on the case and discuss the judgment's approach to personal data and access rights. We also discuss the right to good administration, the right to effective remedies in the Asylum Procedures Directive, and the right to effective judicial protection in the EU Charter. Lastly, we discuss the application of the CJEU judgment by the referring Dutch court in this particular case. Section 5 concludes.[3]

---

[1] CJEU, YS. and M. and S. v. Minister of Immigration, Integration and Asylum (C-141/12 and C-372/12).

[2] Directive 95/46/EC of the European Parliament and the Council of Europe of 23 November 1995 on the protection of individuals with regard to the processing of personal data and on the free movement of such data (OJ 1995, L 281/31).

[3] This comment is based on an article we wrote in Dutch in *Asiel- en Migrantenrecht* 2014, no. 7, p. 299-301. The authors thank Paddy Leerssen for his help.

## 2 Minutes in Dutch asylum procedures

Minutes are drawn up by an official of the Dutch Immigration and Naturalisation Service as an internal opinion on whether to grant a residence permit or not.[4] The minute is part of the preparatory process for adopting the draft decision, but is not part of the final decision. In general, the minute contains information about the applicant, such as his or her name, date of birth, nationality, gender, ethnicity, religion, details of the procedural history, and the statements made by the applicant. In addition, the minute contains a preliminary decision or analysis in which the current rules are applied to the facts of the case. This analysis varies in size from a few sentences to a few pages.

Until 2009, Dutch immigration authorities provided asylum seekers access to the minute, including the legal analysis. According to the Minister of Immigration, Integration and Asylum, this policy led to a significant workload, and asylum seekers often misinterpreted the legal analyses. Therefore, since 14 July 2009, requests for the provision of minutes have been systematically rejected. Instead of a copy of the minute, the applicant now receives an overview of the personal data in the minute, as well as information about the sources of the data and the parties to which the data may have been communicated.

Access to the minutes underlying an asylum decision can be necessary to safeguard the right to effective remedies for the asylum seeker. When appealing against a negative decision, the asylum seeker needs to know the facts and legal grounds upon which a decision is based. The statement of reasons included in a negative decision may provide enough information, but sometimes, important additional information can only be found in the minutes. However, also regarding positive decisions, asylum seekers can have a legitimate interest to gain access to the minutes. This is especially so, as since the adoption of a new Dutch Aliens Act in 2000 in the Netherlands, positive decisions in asylum cases are no longer motivated. According to the Dutch legislator, this rule

[4] See for general information on the Dutch Immigration and Naturalisation Service: <https://ind.nl/en>.

aims to prevent asylum seekers from entering into further proceedings for a stronger status.[5]

For asylum seekers, it can be important to obtain information about the legal reasoning on which a positive decision is based, for several reasons. Firstly, asylum is in theory a temporary status. Therefore, a state may withdraw its protection if changes occur to the situation on which the asylum decision was based. For example, changes in the country of origin (in terms of, for instance, human rights and security) may allow for a revocation of asylum. Secondly, an applicant can have a legitimate interest in access to the grounds of the decision, when the status that is granted offers less protection than the asylum seeker might be entitled to. The Dutch highest administrative court ('Afdeling Bestuursrechtspraak van de Raad van State') consistently rejects appeals in which individuals submit a written request for a written motivation, due to a lack of procedural significance.[6]

During the parliamentary negotiations on the Aliens Act 2000, the Minister of Immigration, Integration and Asylum held that withholding the motivation behind a decision would not limit the procedural rights of the asylum seeker, as he or she was still entitled to ask for access to underlying documents of the decision on the basis of data protection law, even if this would not apply to the minutes of the case.[7] When an asylum seeker's status is withdrawn, Dutch courts may even hold it against the asylum seeker if he or she did not ask access to the files. For example, in 2009, a Libyan national was granted temporary protection because of the overall situation in Libya.[8] During her application, the asylum seeker supplied information on her individual situation to obtain a refugee status. When her temporary status was withdrawn, the Dutch Minister argued that all foreigners must realize that each residence permit is, in principle, temporary, and therefore must take into account that the asylum status may be withdrawn. According to the Minister, the applicant would have no ground for legitimate expectations that her residence permit was granted due to her personal circumstances. To support this lack of legitimate expectations, the Minister argued, undisputed by the Dutch highest administrative court, that the applicant never requested disclosure of the motivation grounds.[9]

---

[5] Parliamentary notes Second Chamber Dutch Parliament, 1999-2000, 26 732, no. 7, p. 4.
[6] See for example, ABRvS 28.09.2011, *JV* 2011/471, ve11002366.
[7] Parliamentary notes Second Chamber Dutch Parliament, 1999-2000, 26 732, no. 7, p. 42-43.
[8] ABRvS 04.08.2014, *JV* 2014, 348, ve14001239, annotation Bem-Reixert.

## 3 The judgment

### 3.1 Background of the cases

In 2009, YS., M. and S. applied for asylum in the Netherlands. The application of YS. was rejected, while M. and S. obtained a temporary status. In separate procedures, the three applicants asked for access to the minutes of their case, based on the Dutch Data Protection Act,[10] the Dutch implementation of the Data Protection Directive. The Minister of Immigration, Integration and Asylum denied these requests.

In both procedures, the courts submitted preliminary questions to the CJEU. The two Dutch courts asked for clarification, firstly, on the interpretation of 'personal data' and the right to have access in Article 12 (a) of the Data Protection Directive and secondly, on the right to good administration in Article 41 of the Charter of Fundamental Rights of the European Union. The CJEU joined the two cases.[11]

### 3.2 Scope of personal data and access rights

Article 12(a) of the Data Protection Directive prescribes that every data subject has a right to obtain 'confirmation as to whether or not data relating to him are being processed and information at least as to the purposes of the processing, the categories of data concerned, and the recipients or categories of recipients to whom the data are disclosed, [and] communication to him in an intelligible form of the data undergoing processing and of any available information as to their source (…)'. To answer the question of whether this right of access applies to the minutes in the asylum procedure, the CJEU deals first with the scope of the concept of 'personal data' of the Data Protection Directive.

---

[9] The Minister held that the applicant could have based this application for motivation on Section 3:48 of the General Administrative Law Act. However, during the aforementioned parliamentary debate on the new Aliens Act, the Minister actually denied the applicability of this provision. Parliamentary notes Second Chamber Dutch Parliament, 1999-2000, 26 732, no. 7, p. 42-43.

[10] See for a non-official translation: <www.coe.int/t/dghl/standardsetting/dataprotection/national%20laws/NL_DP_LAW.pdf>.

[11] The CJEU deals with six questions regarding these three topics.

Personal data is the core concept of the Directive, and its interpretation continues to arouse discussion.[12] The Directive defines personal data, in short, as any information relating to an identified or identifiable natural person.[13] As to whether minutes should be regarded as personal data, the CJEU distinguishes personal data from the legal analysis. According to the CJEU a legal analysis 'is not information relating to the applicant for a residence permit, but at most, in so far as it is not limited to a purely abstract interpretation of the law, is information about the assessment and application by the competent authority of that law to the applicant's situation'.[14] Although the legal analysis found in minutes may include personal data pertaining to the applicant, the analysis does not, according to the CJEU, constitute a form of personal data *per se* for the purposes of Article 2(a) of the Data Protection Directive.[15]

Referring to earlier-case law, the CJEU then gives a narrow interpretation of the goal of the Data Protection Directive, and thus the right of access to personal data protected in Article 12 (a). According to the CJEU, it follows from the *Bavarian Lager* case that Regulation 45/2000 on the processing of personal data by EU institutions 'is not designed to ensure the greatest possible transparency of the decision-making process of the public authorities and to promote good administrative practices by facilitating the exercise of the right of access to documents'.[16] The CJEU says that the Bavarian lager reasoning also applies to the Data Protection Directive. According to the CJEU the purpose of the Data Protection Directive is to protect the right of privacy of the applicant with regard to the processing of data relating to him – and not 'guaranteeing him a right of access to administrative documents'.[17] The CJEU then concludes that only the personal data relating to the applicant for a residence permit contained in the minute 'and, where relevant, the data in the legal analysis contained in the minute are "personal data" within the meaning of Article 2(a) of [the Data Protection Directive], whereas by contrast, that analysis cannot in itself be so classified.'[18] It follows from this reasoning that the right to access of Article 12 (a) to the minutes is limited to that part of the minutes (and its legal analysis) which contains personal data.

---

[12] See for instance Zwenne GJ, *De verwaterde privacywet [Diluted Privacy Law], Inaugural lecture of Prof. Dr. G. J. Zwenne to the office of Professor of Law and the Information Society at the University of Leiden on Friday, 12 April 2013* (Universiteit Leiden 2013); Zuiderveen Borgesius FJ, Improving Privacy Protection in the area of Behavioural Targeting (PhD thesis University of Amsterdam 2014), chapter 5.
[13] Article 2(a) of the Data Protection Directive.
[14] Para. 40.
[15] Para 46.
[16] Para 47.

According to the CJEU, Article 12 (a) leaves it to the Member States to determine how data controllers should respond to access rights, as long as the communication to the data subject is 'intelligible', in other words, as long 'it allows the data subject to become aware of those data and to check that they are accurate and processed in compliance with that directive, so that that person may, where relevant, exercise the rights conferred on him'.[19] The CJEU adds that the data subject does not always have a right to a transcript of the original document. To comply with an access request, the data controller may choose other means 'in so far as the objective pursued by that right of access may be fully satisfied by another form of communication'.[20] According to the CJEU, in order to avoid giving the data subject access to information other than the personal data relating to him, it can be sufficient to grant him or her only a copy of the document or the original file in which that other information has been redacted, or 'a full summary of those data'.[21] Also, considering the underlying case, the CJEU states that 'it is sufficient for the applicant for a residence permit to be provided with a full summary of all of those data in an intelligible form'.[22]

---

[17] Para. 46. See also *Commission v. Bavarian Lager*, C-28/08,
[18] Para 48.
[19] Dictum, sub 2 and para.57.
[20] Para 58.
[21] Dictum, nr 2; see also para 60.
[22] Para. 58-60.

## 3.3 Right to good administration

In the *YS., M. and S.* case, the Dutch courts also submitted preliminary questions on whether asylum applicants could invoke their access rights on the basis of the principle of good administration protected in Article 41 of the EU Charter on Fundamental Rights. This provision offers everyone the right 'to have his or her affairs handled impartially, fairly and within a reasonable time by the institutions, bodies and agencies of the Union'. Pursuant to Article 41 (2), every person has 'to have access to his or her file, while respecting the legitimate interests of confidentiality and of professional and business secrecy'. As the text of Article 41 is not directed at Member States, the question arose whether individuals can invoke this right against their national administrations. In two earlier decisions about national asylum procedures, the CJEU concluded that the principle of good administration is a general principle of EU law, and therefore can be invoked against their national administration.[23]

In *YS., M. and S.*, the CJEU concludes however that the Charter's right to good administration is addressed solely to the institutions, body, offices and agencies of the European Union. Therefore, 'an applicant for a resident permit cannot derive from Article 41 (2) (b) a right to access the national file relating to his application'.[24] Repeating its conclusion in earlier judgments, the CJEU states that the principle of good administration 'reflects a general principle of EU law'.[25] However, as the Dutch judges in this case did not request an explanation of this general principle, the CJEU does not find it necessary to elaborate on the principle.[26]

---

[23] *CJEU, M. v. Ireland* C-277/11, 22 November 2012, para. 83-84 and *CJEU, H.N. v. Minister for Justice, Equality and Law Reform* C-604/12, 8 May 2014, para 50-51. These cases concerned the right to be heard, respectively the requirement of impartiality. See also more recently, *CJEU, Boudjlida* C-249/13, 11 December 2014, in which the CJEU held, in para. 38, that the right to be heard implies motivated decision-making.
[24] Para. 67.
[25] Para. 68.
[26] Para. 68

# 4 Analysis

## 4.1 Limited interpretation of personal data and access rights

The CJEU's reasoning in this judgment on the scope of data protection and access rights can be criticized for several reasons.

Regarding the purpose of access rights in the light of the Data Protection Directive, the CJEU ignores, or even rejects, the more general objectives of data protection law, which by now have been commonly accepted. Data protection law aims not only to protect privacy as the CJEU suggests here, but also, for instance, to mitigate the information asymmetry and power imbalance between data subjects and data controllers, to support good governance, and to enforce the rule of law and transparency of powers.[27] Data protection as an independent, fundamental right has been recognized by the EU legislator in Article 8 of the Charter, alongside the right to privacy. With the Charter the right to have access to personal data obtained the status of a fundamental right: 'Everyone has the right of access to data which has been collected concerning him or her, and the right to have it rectified.' In our view, considering current developments on the use of personal data by private and public organisations and their consequences for individual rights and freedoms, the CJEU expresses an archaic view by limiting the purpose of data protection and access rights to privacy protection.

As noted, the CJEU says that a data controller can comply with a data subject's access right by providing a full summary of the relevant personal data in an intelligible form. The CJEU adds, however, that the data subject must receive the summary of the data in a form that allows him or her, among other things, to exercise his or her data protection rights. Therefore, in many circumstances, a data controller must probably provide a copy of the personal data file – otherwise the data subject cannot exercise his or her rights. Access to minutes is often necessary for the effective enjoyment of the asylum seeker's rights.[28]

---

[27] See amongst others: E. Brouwer, *Digital Borders and Real Rights: Effective Remedies for Third-Country Nationals in the Schengen Information System (Martinus Nijhoff Publishers 2008), p. 200.* See also A.F. Westin, *Privacy and Freedom* (Atheneum 1967), p. 158; P. de Hert & S. Gutwirth, 'Privacy, data protection and law enforcement. Opacity of the individual and transparency of power' in E. Claes, E., A. Duff & S. Gutwirth (eds), *Privacy and the criminal law* (Intersentia 2006).

Our second critical comment concerns the unclear distinction that the CJEU has introduced between (i) personal data in the legal analysis and (ii) the rest of the legal analysis.[29] In practice, dealing with personal files, it will be difficult to distinguish the two categories. Furthermore, the CJEU seems to apply a narrower definition of 'personal data' than the Article 29 Working Party. The Working Party is an advisory body in which national Data Protection Authorities cooperate.[30] While not legally binding, the Working Party's opinions are influential; they give an idea of the views of European national Data Protection Authorities.[31] According to the Working Party, information can 'relate' to a person due to its (i) content, (ii) purpose, or (iii) result. The Working Party says, in short, that information can relate to a person because of a result, when this information can reasonably be expected to have consequences for that person's rights or interests.[32] The Working Party's view implies that the legal analysis in the minutes must be seen as personal data. After all, the legal analysis is applied to determine one's status: will he or she be granted a residence permit, or not?[33]

The CJEU's finding that an analysis in a personal dossier falls outside the concept of personal data can also have consequences outside of asylum law, for instance in the context of the commercial use of personal data. In many modern data information processes, personal data are processed in order to create new personal data which are subsequently added to a personal profile or dossier. For example, behavioural targeting is a type of internet marketing that involves monitoring people's online behaviour, and using the collected information to show people individually targeted advertisements.[34] And credit rating agencies analyse data in order to create risk profiles of individuals. It is important that a data subject can inspect such analyses, exactly because they determine decisions about the data subject.[35]

---

[28] See along similar lines: Jansen M, Arrest HvJ EU inzake begrip persoonsgegevens en karakter inzagerecht, Privacy & Informatie 2014(5), p. 200-206, p. 205-206.

[29] See dictum, under 1.

[30] Article 29 of the Data Protection Directive. See for more details on the Working Party: chapter 4, introduction.

[31] See Gutwirth S & Poullet Y, 'The contribution of the Article 29 Working Party to the construction of a harmonised European data protection system: an illustration of 'reflexive governance'?' in Asinari, V. P. & P. Palazzi (eds), *Défis du Droit à la Protection de la Vie Privée. Challenges of Privacy and Data Protection Law* (Bruylant 2008).

[32] Article 29 Working Party, 'Opinion 4/2007 on the concept of personal data' (WP 136), 20 June 2007, p. 11.

[33] See also Jansen M, Arrest HvJ EU inzake begrip persoonsgegevens en karakter inzagerecht, Privacy & Informatie 2014(5), p. 200-206; Korff D, 'The proposed General Data Protection Regulation: suggested amendments to the definition of personal data', EU Law Analysis 15 October 2014, http://eulawanalysis.blogspot.nl/2014/10/the-proposed-general-data-protection.html.

In our view, the judgment does not imply that data that relate to a person because of a 'result' should generally not be seen as personal data. First, the *YS. and M. and S.* case concerns particular questions regarding asylum procedures, and the CJEU's reasoning cannot be directly applied to other situations. Second, the CJEU has not explicitly rejected the view of the Working Party that information can relate to a person because of its result.

In the *YS., M. and S.* case, the CJEU does not refer at all to Article 8 ECHR, despite the fact that in previous judgments, the CJEU repeatedly underlined that the right to private life as protected in the ECHR is a fundamental principle of Union law. In these cases, the CJEU also referred to the case-law of the European Court of Human Rights (ECtHR) in order to define the scope and content of data protection rights.[36] The CJEU's *YS., M. and S* judgment seems hard to reconcile with the 2009 case *K.H. and others. v. Slovakia*, in which the ECtHR concluded that the right to privacy laid down in Article 8 ECHR may include a positive obligation for states to provide individuals access to the documents of their personal files.[37]

In the Netherlands, disagreement has existed for some time on the question whether access rights include the right to a copy of the documents that contain personal data, or merely a right to a copy of an overview of those data. In the 2006 *Dexia* decision, the

---

[34] See Zuiderveen Borgesius FJ, Improving Privacy Protection in the area of Behavioural Targeting (PhD thesis University of Amsterdam 2014), forthcoming Kluwer 2015.
[35] As Kabel notes, on the basis of registrations credits can be refused, welfare payments withheld and insurance premiums raised (Kabel J, 'Computatoria Locuta, Causa Finita? Consument, Bescherming van Persoonsgegevens en Geautomatiseerde Besluiten', in Van Buren-Dee, J. M., E. H. Hondius & P. A. Kottenhagen-Edzes (eds), *Consument Zonder Grenzen* Deventer: Kluwer,1996.
[36] See CJEU *Rechnungshof v. Österreichischer Rundfunk*, C-465/00, 20 May 2003, para.71 ff. More recently, in *Digital Rights Ireland Ltd.* Joined cases C-293/12 and C-594/12, 4 April 2014, para. 35.
[37] *K.H. and others. v. Slovakia* appl.no. 32881/04, 28 April 2009, para. 47.

Dutch Supreme Court interpreted the access right more extensively than the CJEU.[38] The Supreme Court considered that the Dutch Data Protection Act entitles data subjects to all relevant information. According to the Supreme Court, data controllers in principle cannot refuse the provision of such copies or transcripts in the interest of curbing administrative burdens, since processing large amounts of personal data implies that many data subjects may invoke their access rights.[39] Following this line of reasoning, the Minister's argument for refusing to provide copies of minutes to asylum seekers due to the administrative burden involved, is untenable.

### 4.2 Asylum procedures and effective remedies

The Dutch courts did not ask questions to the CJEU about the Asylum Procedures Directive, or about Article 47 of the EU Charter on Fundamental Rights. In our view, the right to access to the minutes could have been derived from those EU law instruments as well. Pursuant to Article 9 of the former Asylum Procedures Directive (applying at the time of the Dutch procedures), the duty of national authorities to state reasons in fact and law, is limited to negative decisions on asylum applications.[40] According to the same provision, where the applicant is granted a status which offers the same rights and benefits under national and Community law as the refugee status, Member States do not need to state the reasons for not granting refugee status. However, in those cases Article 9 (2) obliges Member State to ensure that the reasons not granting a refugee status are stated in the applicant's file and 'that the applicant has upon request access to his/her file'.

In the recast Asylum Procedures Directive, adopted in 2013 and replacing the former Directive in 2015, the duty to motivate a decision is generally limited to negative decisions. According to Article 11 (2), Member States must in the event of a negative decision provide applicants with information 'in order to clarify the reasons for such decision and explain how it can be challenged' upon request.[41] Applying as well to positive decisions, Article 19 (2) obliges Member States to ensure in procedures at first instance 'that, on request, applicants are provided with legal and procedural information free of charge, including, at least, information on the procedure in the light of the applicant's particular circumstances'. Furthermore, Article 23 (1) of the recast Asylum Procedures Directive obliges Member States to ensure that a legal adviser or other counsellor of an asylum applicant will have access to the information in the applicant's file upon the basis of which a decision 'is or will be made'. Finally, Article 12 (f) on the guarantees for asylum seekers provides that asylum seekers or their counsellors have access to general information the authority has taken into consideration for deciding on their application.[42] Based on these provisions, arguably access to minutes is necessary for the effective enjoyment of the asylum seeker's rights.

---

[38] Dutch Supreme Court, 29 June 2006, ECLI:NL:HR:2007:AZ4663 (Dexia), para. 34.
[39] Our translation.
[40] Council Directive 2005/85/EC Of 1 December 2005 on Minimum Standards on Procedures in Member States for Granting and Withdrawing Refugee Status L 326/13.

This brings us to the relevance of Article 47 of the Charter on the fundamental right to effective judicial protection. In the *YS., M. and S.* case, the Dutch courts did not ask the CJEU about this provision. This right applies whenever a person invokes that his or her right granted under EU law is infringed – also in immigration law procedures covered by EU law.[43] In our view, the right may be invoked whenever applicant can substantiate that his or her right to effective judicial protection is infringed by limiting access to the underlying files of the asylum decision.

As mentioned above, the CJEU does not elaborate on the meaning of the right to access to files as part of the general principle of good administration, because the national courts did not address this matter specifically in their preliminary question. This is disappointing, but at the same time could be considered as advice to lawyers and judges in future asylum cases: lawyers should explicitly invoke the principle, and judges should apply it. Similarly, lawyers should explicitly invoke Article 47 of the Charter on the right to an effective remedy, and the right to informed decision making as provided in the EU Asylum Procedures Directive.

---

[41] Article 11 (2) of the recast-Directive 2013/32 of 26 June 2013, *OJEU* L 180/60, 26.6.2013. This Directive is to be implemented on 20 July 2015.

[42] This includes information of EASO, UNHCR and relevant human right organization dealing with the general situation in the country of origin of country of transit of the applicant and information from experts on particular issues, such as medical, cultural, religious, child-related or gender issues.

[43] CJEU *Mohamad Zakaria*, 17 January 2013, C- 23/12; *M.M. v. Minister for Justice, Equality and Law Reform*, 22 November 2012, C-277/11.

### 4.3 Dutch decisions after CJEU judgment

In December 2014, deciding in two of the three cases leading to the CJEU judgment, the Dutch highest administrative court concluded that the legal analysis in the minutes could not be considered as personal data, and therefore the right of access does not apply to this analysis.[44] Based on the judgment of the CJEU, the Dutch court held that the right to access to personal data does not oblige the authorities to provide a copy of the minutes. However, the Dutch court also underlined that the authorities must provide ‘a full and intelligible’ overview of the personal data contained in the minutes. This means, according to the Dutch court, that it is not sufficient to provide only an enumeration or list of the personal data to the asylum seeker. Furthermore, in one of the two judgments, the highest administrative court dismissed the claim of the immigration authorities that access to the minutes would hamper the ‘free communication’ between officials preparing the decision.[45] The court gives two reasons: first, the authorities are only obliged to give a full and intelligible overview of the personal data. Second, as the court already decided, the legal analysis, which according to the court contains the personal considerations of the official preparing the decision, does not fall within the scope of the right to access. At first sight, it seems a positive development that, in line with the reasoning of the CJEU, the Dutch court obliges national authorities to give a full and intelligible overview of personal data, meaning they are not allowed to provide a mere list of personal data included in the file. However, as result, the Dutch highest administrative court applies the interpretation of the right of access by the CJEU very narrowly.. First, by excluding the legal analysis altogether from the right to access and therefore neglecting the possibility this may contain personal data about the asylum seeker as well. Secondly, by making it explicit that the immigration authorities are not obliged to provide a copy of the personal file, excluding the possibility that with regard to certain documents it could be necessary to provide a copy.

---

[44] Afdeling Bestuursrechtspraak Raad van State, 24 December 2014, ECLI:NL:RVS:2014:4631 and RVS:2014:4626.

[45] ECLI:NL:RVS:2014:4631, para 11.2: “Naar het oordeel van de Afdeling staat het belang van de ongestoorde gedachtewisseling tussen ambtenaren het verstrekken van de [wederpartij] betreffende persoonsgegevens niet in de weg. Daarbij is van belang dat de minister slechts is gehouden een volledig overzicht in begrijpelijke vorm van die persoonsgegevens te verstrekken.”

## 5 Conclusion

In this judgment, the CJEU states that the personal data, such as a name, contained in the legal analysis of a 'minute' document in an asylum procedure are 'personal data' within the meaning of the Data Protection Directive, but that the analysis itself cannot be so classified. In principle this view seems correct, but in practice the distinction may be hard to make. In our view, the CJEU's reasoning about personal data in this judgment should not too quickly be extrapolated to other circumstances, as the judgment concerns specific questions regarding asylum procedures. The judgment does not imply that data that relate to a person because of a 'result' should generally not be considered as personal data.

Applying the CJEU's judgment, a Dutch court concluded that the right of access does not apply to the legal analysis in the minutes, because that legal analysis could not be considered as personal data. This outcome of the CJEU's judgment is disappointing, as, in many circumstances, access to minutes is simply necessary for the effective enjoyment of the asylum seeker's rights.

According to the CJEU, a data controller can comply with a data subject's access right by providing a full summary of the relevant personal data in an intelligible form. It is to be welcomed that the CJEU underlines that the data subject must receive the summary of the data in a form that allows him or her, among other things, to exercise his or her data protection rights. In many cases, a data controller must therefore provide a copy of the personal data file – otherwise the data subject cannot exercise his or her rights.

Finally, the CJEU says that an asylum seeker cannot directly invoke the Charter's right to good administration (Article 41), which includes a right of access to personal files, against national authorities, because Article 41 is addressed solely to the EU institutions. However, in previous judgments dealing with asylum procedures, the CJEU emphasized that the right to good administration laid down in Article 41 of the Charter forms a general principle of EU law. Hence, in our view European law still provides sufficient grounds for effective access rights to the minutes in future asylum cases.

* * *